\documentclass[12pt,a4paper]{article}
\usepackage{graphicx}
\usepackage{amsmath}
\usepackage{times}
\begin{document}

\title{Flux-biased mesoscopic rings}

\author{J. Dajka\footnote{Corresponding author: e-mail: dajka@server.phys.us.edu.pl}, 
\L. Machura, S. Rogozi\'{n}ski, J. \L uczka\\
Institute of Physics, University of Silesia, 40-007 Katowice, Poland}

\maketitle                   

\begin{abstract}
Kinetics of magnetic flux in a thin mesoscopic ring biased by a strong external magnetic field is described equivalently by dynamics of a Brownian particle in a tilted washboard potential.
The 'flux velocity', i.e. the averaged time derivative of the total magnetic flux in the ring, is a candidate for a novel characteristics of mesoscopic rings. Its global properties   reflect the possibility of accommodating persistent currents in the ring. 
\end{abstract}

\section{Mesoscopic rings: two-fluid model}
Mesoscopic devices have attracted much theoretical and practical attention because they are promising for implementation 
in ultra-small hybrid elements to test quantum information theory \cite{makh}. A large class of such devices is 
based on  ring structures, i.e. the Aharonov-Bohm topology. Such a class contains both superconducting (SQUIDs) and non-superconducting devices. 

In this paper we study selected kinetic aspects of persistent currents which can be observed  in normal metal,
semiconducting rings or cylinders and, as probably the most famous examples, in carbon nanotubes or nanotori.   
We focus our attention on  kinetics of magnetic flux in the presence of a strong external static magnetic field. 
We show that it can be modeled in the same way as the dynamics of a Brownian particle moving in a biased washboard potential. 
Here, the analog of the position of the Brownian particle is a total magnetic flux.    
We show that the time derivative of the magnetic flux, i.e. the flux velocity (if we recall 
the analogy to the dynamics of the Brownian particle) depends strongly on the ability of accommodation of persistent currents by the ring. 

Persistent currents are equilibrium currents flowing in the Aharonov-Bohm  systems which are small enough to preserve phase coherence of electrons  \cite{pc,exper}. In ideal samples at the vanishing temperature $T=0$, all electrons are the carriers of such a current. It is not the case at non-zero temperatures $T>0$, when some of the electrons are no longer coherent  and are a source of the 'normal' Ohmic current. 
Let us consider now a  mesoscopic ring placed  in a uniform 
magnetic field
$B$ in the 3-dimensional  space. Because of the self-inductance $L$,
the electric current $I$ will induce a magnetic flux $\phi$ in
the ring. Therefore, the flux and the current in the
ring are coupled according to the  expression
\begin{equation}
\phi= \phi_{e}+L I = \phi_{e}+L[I_{coh}+I_{dis}].
\end{equation}
The flux $\phi_{e}$ is induced by
the external  magnetic field $B$.
The total current $I$ is a sum of the coherent current $I_{coh}$ and the Ohmic dissipative current $I_{dis}$. The coherent current is assumed to be a linear combination of the paramagnetic and diamagnetic contributions. This is 
related to occurrence, with a probability $p$, of the so called current channel with an even number of coherent electrons or 
 an odd number of coherent electrons, with a probability $1-p$.  Hence, with $\phi_0=h/e$, it reads \cite{cheng} 
\begin{eqnarray}
I_{coh} = I_{coh}(\phi,T)=
I_0\sum_{n=1}^{\infty}A_n(T/T^*)\left\{ p \sin\left(2\pi n\frac{\phi}{\phi_0}
\right) 
+ (1-p) \sin\left[2\pi n\left(\frac{\phi}{\phi_0} +\frac{1}{2}\right)
\right]\right\}.
\end{eqnarray}
 The amplitudes take the form \cite{cheng}
\begin{equation} \label{amp}
A_n(T/T^*)=\frac{4T}{\pi
T^*}\frac{\exp(-nT/T^*)}{1-\exp(-2nT/T^*)}\cos(nk_Fl_x)\;.
\end{equation}
The characteristic temperature $T^*$ is determined from  the
relation $k_BT^*=\Delta_F/2\pi^2$, where $\Delta_F$ marks the energy
gap,  $k_F$ is the momentum at the Fermi surface and 
$l_x$ is the circumference of the ring. 
The parameter  $I_0$ is the maximal value of the persistent current 
at temperature $T= 0$. 

 The dissipative current $I_{dis}$ is determined by
 the Ohm's law and Lenz's rule \cite{lucz1},
\begin{equation}\label{inor}
I_{dis} = I_{dis}(\phi, T)=-\frac{1}{R}\frac{d\phi}{dt}
+\sqrt{\frac{2k_BT}{R}}~\Gamma(t)\;,
\end{equation}
where $R$ is resistance of the ring, $k_B$
denotes the Boltzmann constant and $\Gamma(t)$  describes 
thermal, Johnson-Nyquist fluctuations of the Ohmic current. This
thermal noise is modeled by 
the Gaussian white noise  of  zero average, i.e.,
$\langle \Gamma(t)\rangle=0$ and  $\delta$-auto-correlation function
$\langle\Gamma(t)\Gamma(s)\rangle=\delta(t-s)$. The
noise  intensity $D_0=\sqrt{2k_BT/R}$ is chosen in accordance with
the classical fluctuation-dissipation theorem.  

From equations (1)-(4),   we get  the Langevin equation 
governing the dynamics of the magnetic flux \cite{my1}:
\begin{equation}\label{lan1}
\frac{1}{R}\frac{d\phi}{dt}=-\frac{1}{L}(\phi-\phi_{e})+I_{coh}(\phi,T)+\sqrt{\frac{2k_BT}{R}}\Gamma(t).
\end{equation}
It can be rewritten in the form 
\begin{equation}\label{lan2}
\frac{1}{R}\frac{d\phi}{dt}=-\frac{dW(\phi)}{d\phi}
+\sqrt{\frac{2k_BT}{R}}\Gamma(t),
\end{equation}
where the generalized potential $W(\phi)$ reads 
\begin{eqnarray}\label{W}
W(\phi)&=&  \frac{1}{2L} \left(\phi^2 - 2\phi_e \phi\right) \nonumber\\
&+&\phi_0 I_0\sum_{n=1}^{\infty}
\frac{A_n(T/T^*)}{2\pi n} \left\{ 
p\cos\left(2\pi n \frac{\phi}{\phi_0}\right)+(1-p)\cos\left[2\pi n\left(
\frac{\phi}{\phi_0}+1/2\right)\right]\right\}.
\end{eqnarray}
Eq. (\ref{lan2})  has been  analyzed under various regimes \cite{my1}. 
In the following, we study specific regime of this  system. 
\begin{figure}[htb]
\begin{minipage}[t]{.45\textwidth}
\includegraphics[width=\textwidth]{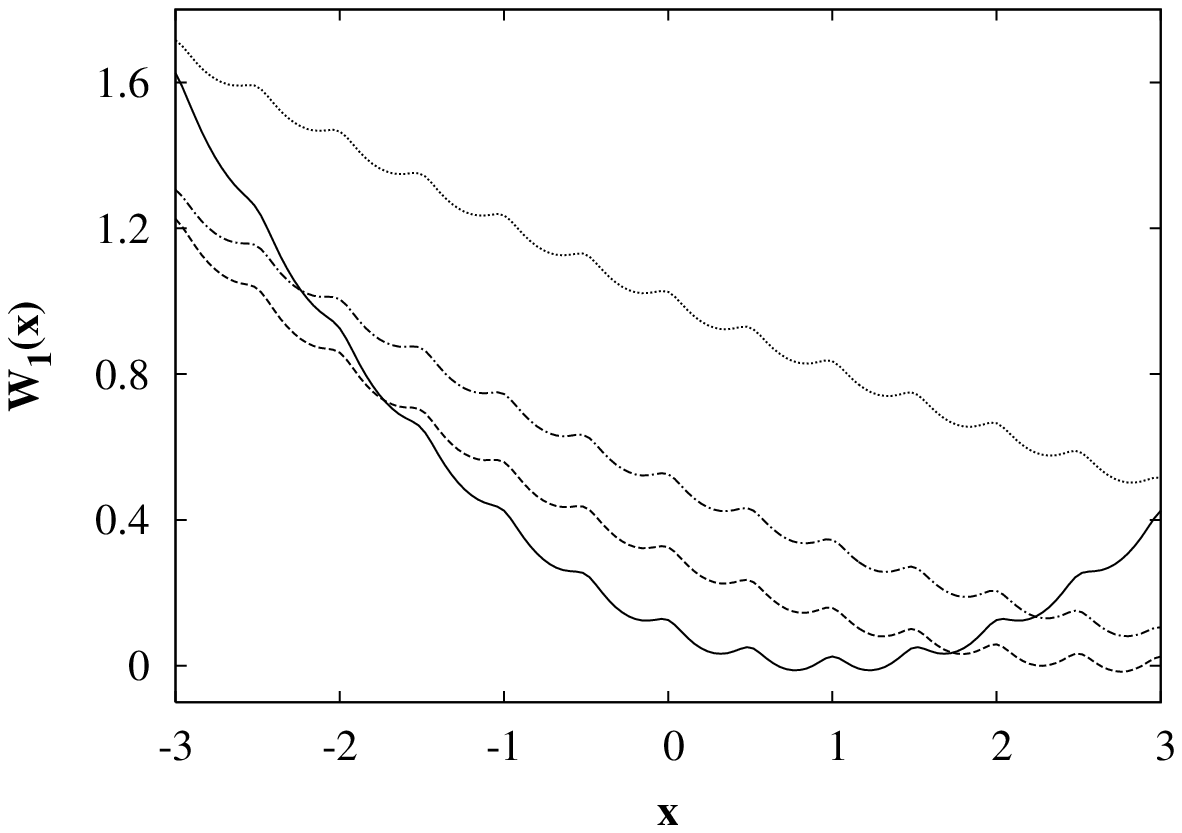}
\caption{The rescaled generalized potential (\ref{W1}) for  $p=1/2$ 
and temperature  $T_0=T/T^*=0.5$. The potential approaches the washboard 
 potential  (\ref{V}) when the  external flux $x_e$ increases and the ratio 
$x_e/L_1 = \phi_{e}/(I_0 L)$ is fixed. This ratio is $1/5$ and 
$x_e=1, L_1=5$ (solid line), $x_e=3, L_1=15$ (dashed line), $x_e=5, L_1=25$ 
(dash-dotted line) and  $x_e=20, L_1=100$ (dotted line).} 
\label{fig:1}
\end{minipage}
\hfil
\begin{minipage}[t]{.45\textwidth}
\includegraphics[width=\textwidth]{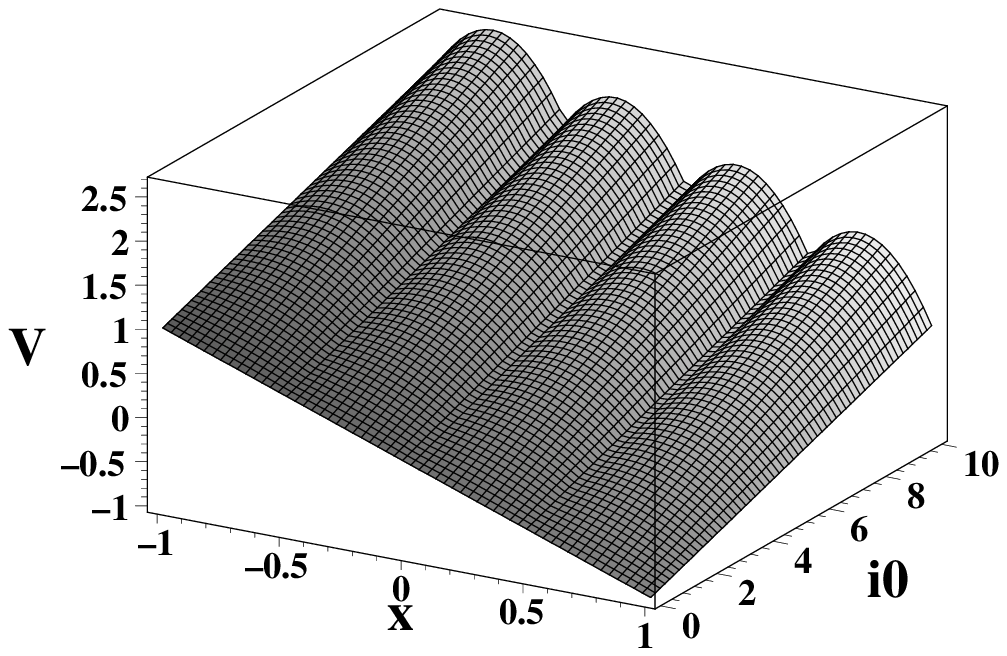}
\caption{The washboard potential $V(x)$  defined by Eq. (\ref{V}) for 
 $p=1/2$ and temperature $T_0=0.5$. For the rescaled current 
amplitude $i_0>i_c$, 
the potential barriers appear. The critical value  of the current 
amplitude $i_c$ is defined by the conditions $V'(x)=0$  and $V''(x)=0$. 
For the presented set of parameters $i_c \approx 2$.}
\label{fig:2}
\end{minipage}
\end{figure}

\section{Flux-biased regime}

We intend to investigate the flux-biased regime which is defined in the 
following way \cite{devoret}: Let the external magnetic 
field $B$ increases giving rise to  increase of the magnetic 
flux $\phi_{e}$. 
Let additionally the self-inductance  $L$ increases. 
Formally, we perform the limit 
$\phi_{e}\to\infty$ and $L\to\infty$ in such a way that the ratio 
$\phi_{e}/L=I_e$ is fixed. 
In this limit, the generalized potential $W(\phi)$ approaches 
a washboard form. Indeed, in 
  Fig. 1, we present four forms of the dimensionless generalized 
potential 
\begin{eqnarray}\label{W1}
W_1(x)&\equiv&  \frac{W(\phi)}{\phi_0 I_0}=  \frac{1}{2L_1} 
\left(x^2-2x_e x\right) \nonumber\\
&+& \sum_{n=1}^{\infty}
\frac{A_n(T/T^*)}{2\pi n} \left\{ 
p\cos\left(2\pi n x \right)+(1-p)\cos\left[2\pi n\left(
x+1/2\right)\right]\right\}, 
\end{eqnarray}
with the dimensionless 
flux  $x=\phi/\phi_0$, $x_e=\phi_e/\phi_0$ and the dimensionless
inductance $L_1=LI_0/\phi_0$.  
We can notice that for the fixed ratio $x_e/L_1=\phi_e/(LI_0) = I_e/I_0 = 1/5$ and for $x_e=20$ and $L_1=100$, the potential $W_1(x)$  is very 
 well approximated 
by the biased washboard potential for large (but finite) number of periods of the 
coherent current $I_{coh}(\phi, T)$.
In the flux-biased regime,   the Langevin equation (5) 
takes the form  
\begin{equation} \label{lan3}
\frac{dx}{d\tau} = -\frac{dV(x)}{dx}+\sqrt{2D}\xi(\tau),
\end{equation}
where  the dimensionless time 
$\tau = t/\tau_0$ with $\tau_0=\phi_0/RI_e$ and the biased washboard potential 
$V(x)$  (see Fig. 2) reads
\begin{eqnarray}\label{V}
V(x)&=& V(x, T_0) = \frac{L_1}{x_e}W_1(x) \nonumber\\
&= &-x+i_0\sum_{n=1}^{\infty}\frac{A_n(T_0)}{2\pi n} \left\{ 
p\cos(2\pi n x)+(1-p)\cos\left[2\pi n(x+1/2)\right]\right\}
\end{eqnarray}
with  the rescaled current amplitude $i_0=I_0/I_e$. The rescaled 
zero-mean Gaussian white noise 
$\xi(\tau)$ has the same $\delta$-auto-correlation function as $\Gamma(t)$. 
Its intensity $D=k_0T_0$, where the dimensionless temperature $T_0=T/T^*$ and 
$k_0=k_BT^*/\phi_0I_e$ is the ratio of thermal energy at the characteristic 
temperature $T^*$ to the energy of the flowing current $I_e$ induced by 
the elementary flux $\phi_0$.   

The Langevin equation (\ref{lan3}) can be interpreted in terms of the 
overdamped motion of the Brownian particle in the washboard potential 
(\ref{V}).  
The periodic part of this  potential  is a 'ratchet type' potential 
\cite{lucz},  
i.e.   $V(x)$ does not posses the reflection symmetry.  
 Let us notice that for $p=1/2$,   there is an  additional periodicity    
$V'(x+1/2)=V'(x)$ presented in the system.

\begin{figure}[htb]
\begin{minipage}[t]{.45\textwidth}
\includegraphics[width=\textwidth]{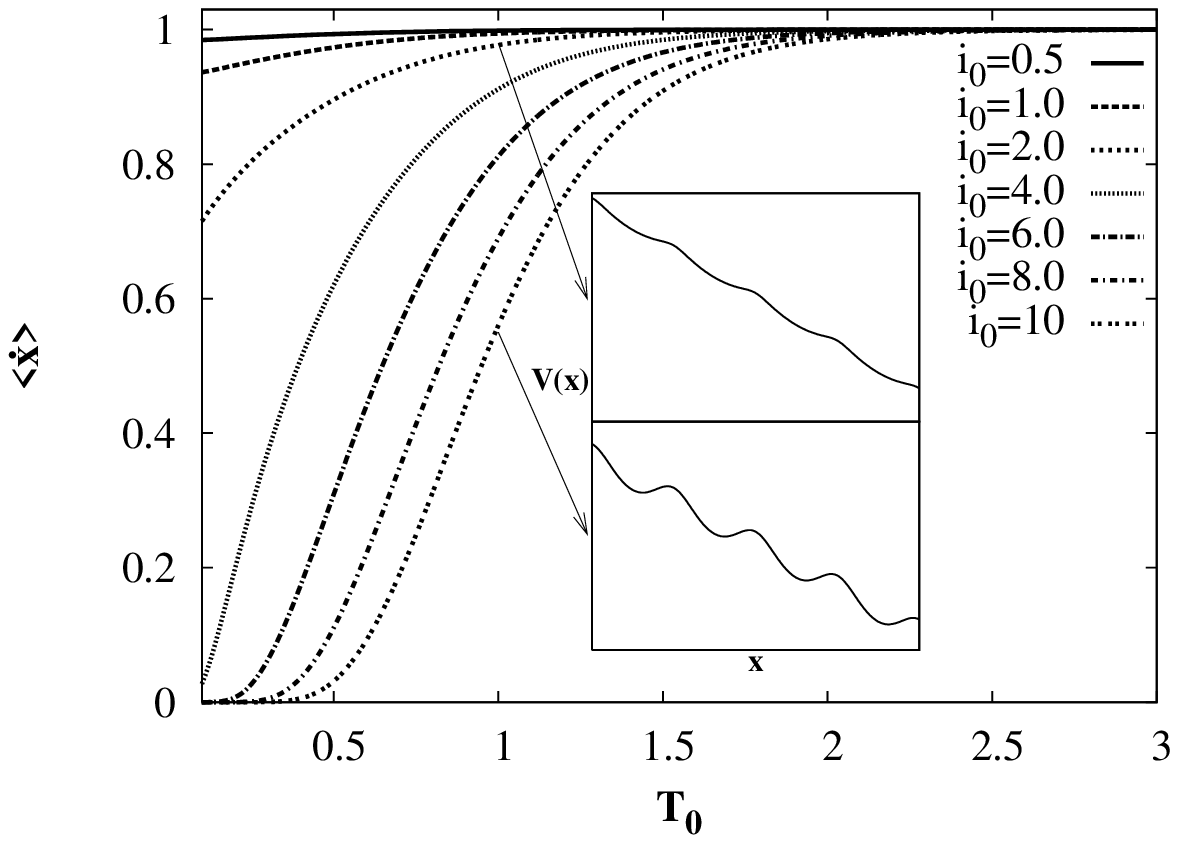}
\caption{The averaged stationary flux velocity as a function 
of temperature $T_0$ for 
several (both  subcritical and supercritical) values   
of $i_0$ and fixed $p=1/2$.   Inset: The corresponding  washboard potentials 
(\ref{V}) at $T_0=1$ are depicted for $i_0=2$ (upper panel) and $i_0=10$ (lower panel). }
\label{fig:3}
\end{minipage}
\hfil
\begin{minipage}[t]{.45\textwidth}
\includegraphics[width=\textwidth]{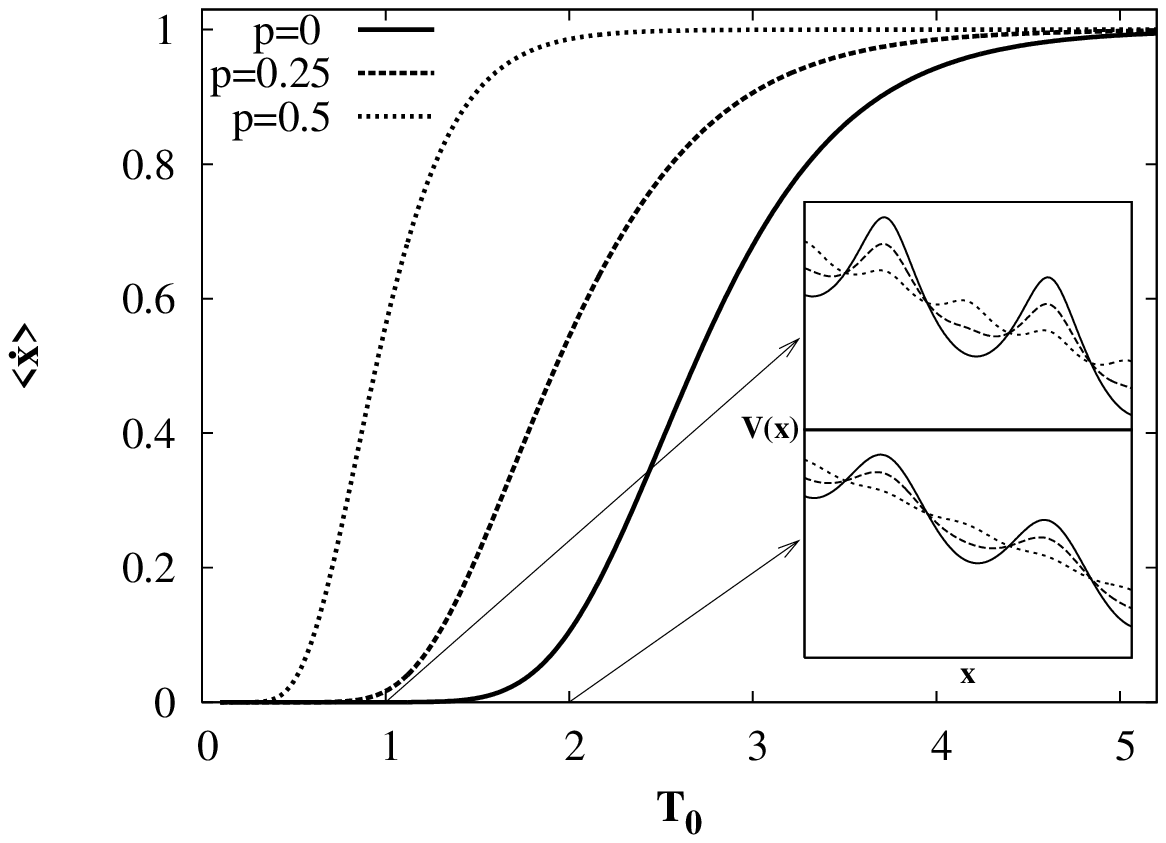}
\caption{The averaged stationary flux velocity as a function of $T_0$ 
for three  values of $p$ and  fixed $i_0=10$. Inset: The corresponding 
washboard  potentials (\ref{V}) 
 at  $T_0=1$ (upper panel) and $T_0=2$ (lower panel) are presented for three values of $p=0, 0.25, 0.5$.}
\label{fig:4}
\end{minipage}
\end{figure}

\section{Flux velocity}

In this paper we shall study the averaged (with respect to the noise realizations)  stationary flux velocity $\langle\dot{x}\rangle$ which is 
given by the formula \cite{ratchet}:
\begin{eqnarray}
\langle \dot{x}\rangle= N\left[1-\exp\left(\frac{V(1,T_0)-V(0,T_0)}{D}
\right)\right],\\
N^{-1}= D^{-1}\;\int_{0}^{1}dx\int_{x}^{x+1}dy\exp\left(\frac{V(y,T_0)-V(x,T_0)}{D}
\right).
\end{eqnarray}

The flux velocity is a function of the system  parameters. The first parameter is the
temperature. The second one is the current amplitude $i_0$, which reflects the ability 
of accommodating persistent currents.  
The third one, $p\in [0,1]$, describes the structure of current channels as it 
is the probability of occurring current channel carrying even number of phase 
coherent electrons. Let us notice that, due to quantum size effects,  persistent 
currents are always present in a sufficiently small system, i.e. there are 
always electrons maintaining their phase coherence when moving around the 
ring. The problem is if $i_0$ is sufficiently large for those electrons to 
produce significant contribution to the total current flowing in the system 
at a given temperature. Numerical results show that upon inspection of the properties of 
the flux velocity $\langle \dot{x}\rangle$ one can infer when the given ring 
is able to accommodate persistent current of a significant 
amplitude at a given temperature.   
In Fig. \ref{fig:3} we present the relation between the flux velocity and the temperature $T_0$ for several different values of the current amplitude $i_0$.
The general tendency is that in the presence of persistent currents 
the flux velocity is suppressed at the low temperature  $T_0$ or large $i_0$. 
There are two classes of the systems split by the 
critical value $i_c$ of the current amplitude $i_0$, which determines the inflection points of the potential.
This qualitative change is defined by a set of two equations: 
$V'(x)=0$ and $V''(x)=0$ and is presented in Fig \ref{fig:2}. 
For $i_0 > i_c$,  barriers of the 
potential $V(x)$ exist.  For $i_0 < i_c$, the potential is a monotonic 
function of the flux $x$. 
Systems from the first class, with the supercritical amplitude $i_0>i_c$, 
exhibit {\it vanishing} flux velocity 
for $T_0\rightarrow 0$. 
For the second class systems, with the subcritical 
amplitude $i_0<i_c$, the flux velocity decreases but remains {\it finite} 
as temperature  $T_0 \to 0$. It is clear that in the formal limit $i_0=0$, 
the flux 
velocity is constant,  $\langle \dot{x}\rangle_{|i_0=0}=1$. The critical 
value $i_c\approx 2$ is estimated for the ring with $p=1/2$.

In Fig. 4,  the probability $p$ of a channel with an even number of coherent electrons is chosen as a parameter.  For the system with statistically equal number of channels of both types ($p=1/2$), the flux velocity is greater than for systems with one type of channels dominating over the other (e.g. $p=0$). The results obtained for $p=1/4$ and $p=0$ coincide with $p=3/4$ and $p=1$ respectively and there is no way to distinguish with type of channels dominates in the ring.   

In order to quantify the effect, one can define the 'susceptibility', i.e. the temperature derivative of the averaged stationary flux velocity at fixed values of other parameters. 
Its monotonicity characterizes the possibility of obtaining persistent currents in the system. 
With this function one can associate  a measure of the ability of accommodating persistent currents in the ring. This measure could be defined as a distance, in the sense of a metric in a function space, between the given (non-zero) 
susceptibility   and the  zero susceptibility.    
 
In conclusion, we have shown that performing suitable limiting procedure 
 one can obtain new significant  informations about persistent currents in 
 mesorings. 
Investigations of  global properties of the flux velocity can serve as an additional
 characteristics of mesoscopic rings.  The perfect example is the monotonicity
of the temperature derivative of the flux velocity  or its asymptotic  behavior at low temperatures which  carries  information about possibility  of appearing persistent currents in the ring.

\section*{Acknowledgments}
Work supported by the Polish Ministry of Science and Higher Education 
under the grant N 202 131 32/3786. 
  J. D. acknowledges the support of the Foundation for Polish Science under grant SN-302-934.

\end{document}